\documentclass[11pt]{article}
\usepackage{moriond,epsfig}

\def\Journal#1#2#3#4{{#1} {\bf #2}, #3 (#4)}

\def\AA{{\em Astron. \& Astroph.}}

\def\be{\begin{equation}}
\def\ee{\end{equation}}
\def\bea{\begin{eqnarray}}
\def\eea{\end{eqnarray}}

\begin{document}
\vspace*{4cm}
\title{SNIa photometric studies in SNLS}

\author{ N. PALANQUE-DELABROUILLE \\
(on behalf of the SNLS experiment)}

\address{Bat. 141, IRFU-SPP, CEA Centre de Saclay, 91191 Gif sur Yvette, Cedex, France}

\maketitle\abstracts{
The discovery of accelerated expansion using supernova surveys has been one of the most surprising discoveries in cosmology in the past ten years. Present and future surveys, among which SNLS, JDEM or LSST, are based on samples of a few hundreds to a million supernovae. The measurement of their spectroscopic redshifts to investigate dark energy properties is already by far the limiting aspect of such surveys. In this paper, I will discuss and illustrate with SNLS data an approach based solely on photometry to both select supernova candidates and determine their redshift.}

\section{Introduction}

From 2003 to 2008, the Supernova Legacy Survey (SNLS) collected data with the MegaCam imager, a 1 square degree array of 36 CCD with a total of 340 million pixels, over four 1-square degree fields. The data were obtained in a rolling search mode, with a typical time sampling of one point every three to four nights, in four different visible frequency bands $g_M$, $r_M$, $i_M$ and $z_M$. The instrument and scanning strategy were designed specifically for the detection of Type Ia supernovae (SNIa) in the redshift range between $0.2$ and $1.0$. 

The standard SNIa selection for the cosmological analyses  in SNLS~\cite{A06} are based on real-time detection and follow-up spectroscopy (thereafter RTA for Real Time Analysis). Despite the significant amount of time allocated to SNLS for the spectroscopy of its candidates, a spectrum could be obtained for roughly half of them only. This justified additional studies based solely on photometry.  The SNLS photometric analyses described here are independent of the RTA, based on different data processing and selection. As such, they can provide a cross-check to estimate possible biases of the standard scenario. They can also be used for studies requiring larger sets of SNIa, as well as for the selection of other transients. 
For instance, a photometric analysis of the SNLS data has already led to the identification of a sample of 117 core-collapse supernovae (SNCC), from which was derived the most precise measurement to-date of the rate of SNCC at a mean redshift $z\sim 0.3$ (Bazin et al., 2009~\cite{bazin09}).     

These proceedings present two major steps towards the use of SNIa from pure photometric studies, i.e. without requiring any spectroscopic information, to derive cosmological constraints. The photometric identification of SNIa, summarizing work which can be found in Bazin et al. (2010) in prep., is described in section~\ref{sec:identification}. The determination of the photometric redshift of SNIa (see Palanque-Delabrouille et al., 2010~\cite{palanque} for details) is presented in section~\ref{sec:redshift}.

\section{SNIa photometric identification}\label{sec:identification}

\subsection{Detection catalog}

Image subtraction was used to search for the appearance of transient events in the 3-year SNLS data (from 2003 to 2006) and to measure their light curves. Because most of the signal from SNIa in the redshift range $[0.2; 1.0]$ is expected to lie in the $i_M$ band, this is where the catalog of detections was built, leading to $\sim$ 300,000 events.  

These were dominated mostly by saturated signals from bright objects which were not perfectly subtracted, by a large contribution from AGNs and a lesser one from variable stars.  Unlike supernovae which are expected to have a flat light-curve before the explosion or a year after, AGNs and long-term variable stars usually exhibit variations over several years. Pollution of the measurements from saturated stars also produces random signals over the entire light curve. Most of the background detections were thus rejected by criteria based on the  simultaneity of the signal between the various filters, and on considerations of light curve stability outside the time range of the main fluctuation. This led to a catalog of about 1500 detections. 

To proceed with the selection of SNIa, a redshift information is required. The detections were therefore matched with galaxies from an external catalog of galactic photometric redshifts~\cite{ilbert} obtained from stacked images of the same fields. Whenever possible (successful match between the event and a unique host galaxy, and available photometric redshift for the galaxy) the events were assigned the photometric redshift $z_{\rm gal}$ of their host galaxy. This reduced the catalog to $\sim 1200$ detections with an assigned photometric redshift. 

\subsection{SNIa selection}

The SALT2~\cite{guySALT2} light curve fitter was applied simultaneously over the four-band light curves of each event to provide an estimate of the colour $C$, the stretch $X_1$ and the rest-frame B band peak magnitude $m_B$, under the assumption that the event was a SNIa. These characteristics were used to further select SNIa and discriminate against two major contaminations:\\ 
-  core-collapse supernovae,\\
- SNIa which were accidentally assigned a photometric redshift $z_{\rm gal}$ significantly different from their actual reshift. \\
The selection criteria were set up using synthetic SNIa, synthetic SNCC, and the subset of our events which happened to have also been selected in the RTA. The latter then have a spectroscopic redshift available ($z_{\rm spe}$) and a confirmed type (Ia or CC). 

The major steps of the analysis are described below.

Long duration events ($X_1>4$) were rejected because almost none of the SNIa fall in that category, in contrast to almost all the plateau core-collapse and about 10\% of the other SNCC.  

\begin{figure} [h]
\begin{minipage}{.5\textwidth} \centering
\epsfig{figure=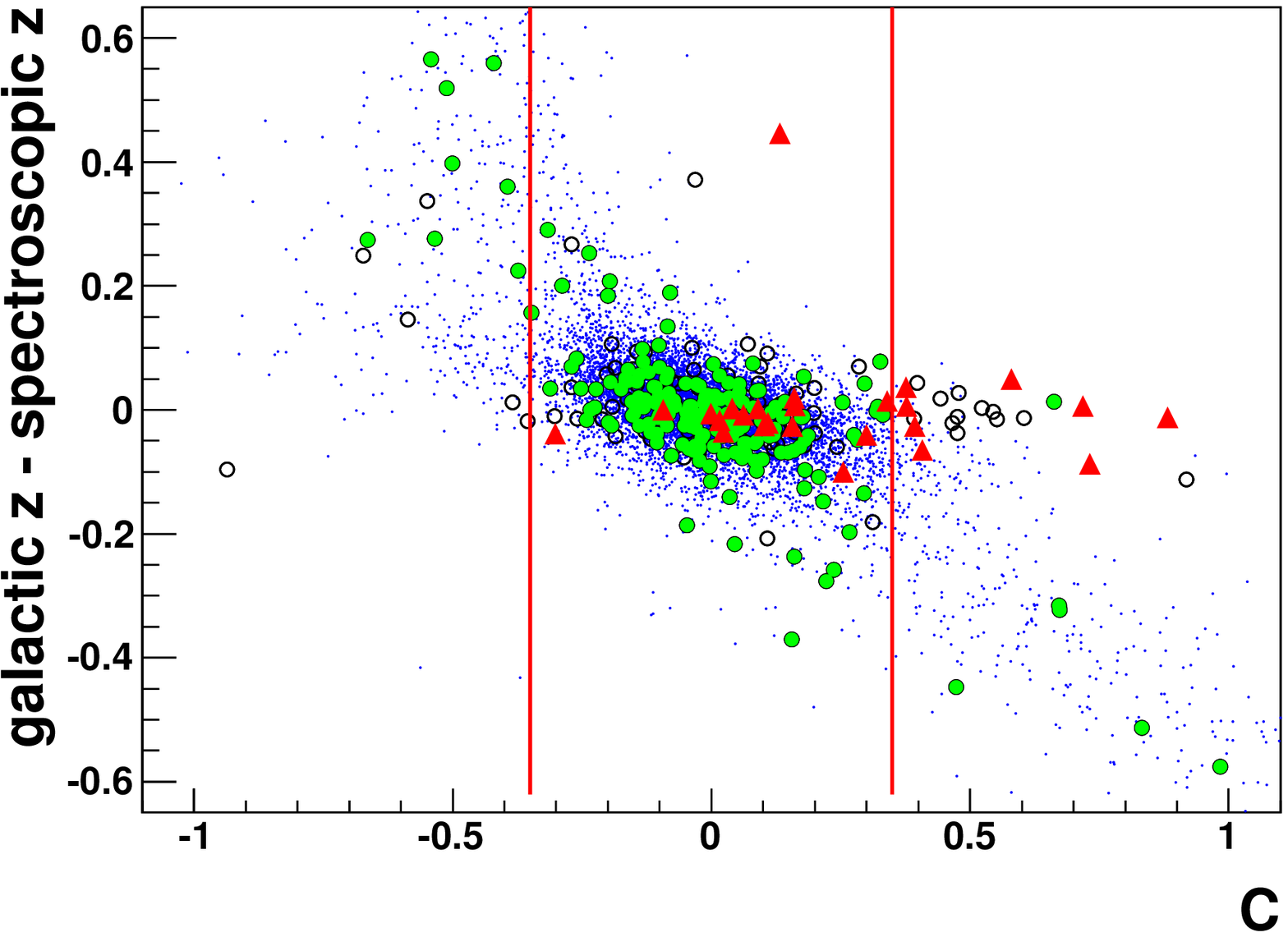,width = \textwidth}
\end{minipage} \hfill \begin{minipage}{.5\textwidth}\centering
\epsfig{figure=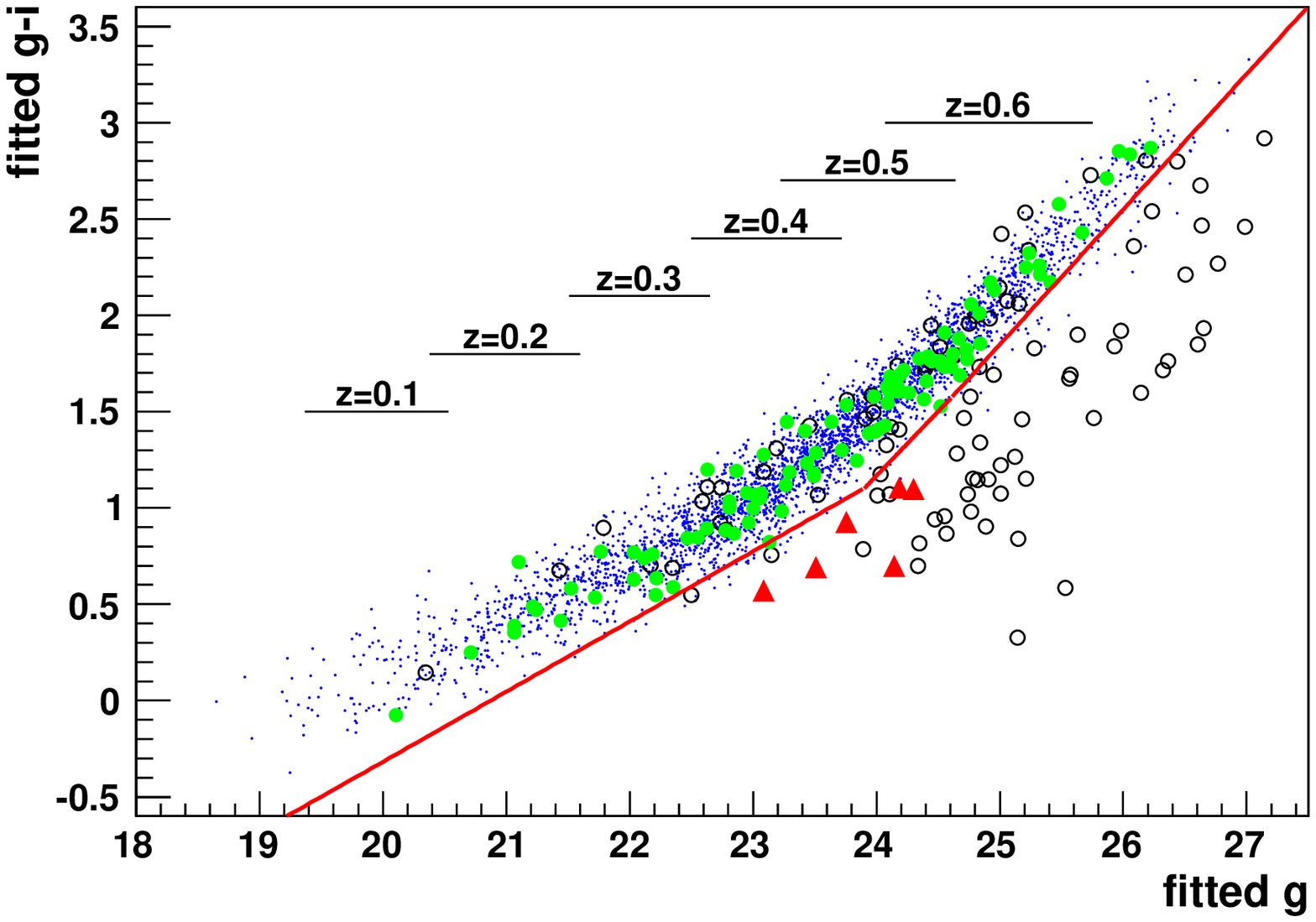, width = \textwidth}
\end{minipage}
\caption{Left: Difference in redshift assignments $z_{\rm gal} - z_{\rm spe}$ as a function of color $C$. Extreme values are rejected. Right:  $g-i$ vs. $g$ color-magnitude diagram. Events below the red curve are rejected. In both plots, synthetic SNIa are in dotted blue, data in black circles. Green filled circles stand for data events spectroscopically confirmed as SNIa, and red triangles for spectroscopically confirmed SNCC.
\label{fig:color}}
\end{figure}

As illustrated in figure~\ref{fig:color} (left plot), rejecting extreme colors (requiring $|C|< 0.35$) was very efficient against both types of contaminants. While SNIa have $\langle C \rangle = 0$ with a r.m.s. of 0.1, SNCC exhibit instead $\langle C \rangle = 0.3$. The constrain on color thus rejected 40\% of the remaining SNCC. In addition, a strong correlation was observed between events fitted with extreme color values and events with a bad redshift assignment: about 70\% of the synthetic SNIa with $|z_{\rm gal} - z_{\rm spe}| > 0.2$ were removed by the above constraint on color.

The final major step to purify the photometric sample of SNIa was based on color magnitude diagrams, as the one illustrated in figure~\ref{fig:color} (right plot). In these diagrams, SNIa populate a thin band while SNCC lie in a broad region which is shifted w.r.t. the SNIa band.

The photometric analysis briefly summarized above selected a total of 485 SNIa. A fraction of these (175 SNIa, thereafter the ``identified" sample) had also been selected in the RTA pipeline; they were confirmed as Type Ia from spectroscopy, and their redshift was measured at the same time. They can therefore be used as a test sample to compare their characteristics with those of the additional SNIa solely selected from photometry (310 SNIa) and thereafter called the ``unidentified" sample. The main difference between the two subsamples lies in their magnitude distribution. Because in SNLS a spectrum could only be obtained for events brighter than $i\sim 23.4$,  the unidentified sample extends about 1 magnitude deeper than the identified one. This translates into an average redshift of 0.6 (resp. 0.9) for the identified (resp. unidentified) sample. Despite this difference, both subsamples exhibit similar properties. For instance, when limited to a common range of bright events ($i<23$), both subsamples exhibit similar dependences of the residuals to the Hubble diagram with color or with stretch (the so-called ``brighter-bluer'' and ``brighter-slower'' relations).

This photometric selection increased by about a factor of 2 the set of SNIa found in the SNLS data. These can already allow new studies where additional statistics is necessary (e.g. to split the events into subsamples).  From simulated SNCC, the contamination of this photometric sample was estimated to be  $\sim 0$ from plateau SNCC and of $\sim 3\%$ from other SNCC. 

\section{SNIa photometric redshift}\label{sec:redshift}

The above analysis was using the host galaxy photometric redshift as the assigned supernova redshift. We saw however that the external catalog can be incomplete (resulting in a 17\% loss in our case). We can also assign the wrong host (resulting in an arbitrary supernova redshift). Even when all goes well, the uncertainty on a galaxy photometric redshift is at present only of order 5\%. It is thus reasonable to assume that using the time-dependent information available in supernovae light curves could result in a redshift with smaller uncertainties. 
The derivation of a supernova redshift from its light curve is the aim of the work presented in this section.

Because, by definition, the redshift of the supernova shifts its spectrum towards larger wavelengths, the amount of flux measured in the four bands of SNLS is also affected. The main information about redshift is thus contained in the observed colors of a SNIa. The measurement of the time-evolution of these colors, through the use of the full multi-band light curves, is crucial to break inherent color-redshift degeneracies. 

Recently, both the SNLS~\cite{palanque} and the SDSS~\cite{kessler} experiments have developed methods to estimate supernova redshifts using light curve fitters, where the redshift is determined as any of the other free parameters. One of the main difficulties in such a procedure is the mandatory  initialization of the parameters, to avoid falling in an arbitrary local minimum. The initialization  is usually done in several steps. In SNLS, the first step is a scan of the fit $\chi^2$ as a function of redshift, where both color $C$ and stretch $X_1$ are fixed to 0. This yields an initial estimate of the redshift used thereafter. A second fit, with a Gaussian prior set on color, is done to determine the stretch value. The last fit, with redshift and stretch constrained to the result of the second step, lets the color unconstrained in order to estimate its value.

\begin{figure} [h]
\centering
\epsfig{figure=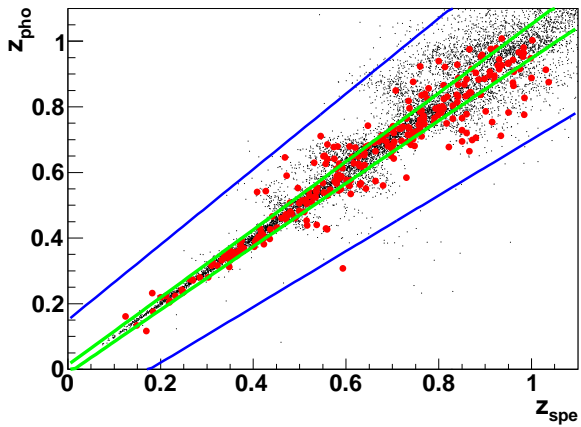,width = .53\textwidth}
\caption{Photometric ($z_{\rm pho}$) vs. spectroscopic ($z_{\rm spe}$) redshift of SNIa. The green lines are for $z_{\rm pho} =z_{\rm spe} \pm 0.022(1+z_{\rm spe})$, representing the average precision attained up to $z=1$, and the blue lines for $z_{\rm pho}= z_{\rm spe} \pm 0.15(1+z_{\rm spe})$ to visualize the catastrophic redshifts. Red circles for data SNIa, black points for simulated ones.
\label{fig:redshift}}
\end{figure}

The performance of the method is estimated using the SNIa (whether spectroscopically confirmed or selected with the photometric analysis of section~\ref{sec:identification}) which have their redshift (or that of their host) known from spectroscopy. The results are illustrated in figure~\ref{fig:redshift}. 

The resolution was defined as $\sigma_{\Delta z/(1+z)}  \equiv 1.48\times{\rm median}[|\Delta z|/(1+z)]$ (a robust estimate of the r.m.s.), and the outlier rate $\eta$,  or rate of catastrophic errors, 
as the proportion of events with $|\Delta z|/(1+z)>0.15$. With the method summarized here, we obtained $\sigma_{\Delta z/(1+z)} = 0.022$ on average over the entire redshift range and $\eta = 0.7\%$. This is a significant improvement over the current performance of galactic photometric redshifts, where typically $\sigma_{\Delta z/(1+z)} = 0.037$ and $\eta = 5.5\%$.
\\

SNIa photometric redshifts of this precision will be useful for future experiments (such as the Large Synoptic Survey, the Dark Energy Survey or Pan-STARRS) which aim to discover up to millions of Type Ia supernovae but without spectroscopy for most of them. 

One should keep in mind,  however, that it might remain useful to use the SNIa photometric redshift as an improvement for an already selected sample of SNIa (based for instance on a redshift coming from the host galaxy, as described here). The reason is that an analysis combining both the identification of SNIa and their redshift determination from the SNIa light curves only (without spectroscopy nor using an external catalog of host galaxy photometric redshift) might result in a larger  contamination from SNCC for instance. Such a combined analysis will be the topic of a future study.

\end{document}